\documentclass[11pt,a4paper]{article}
\usepackage{jheppub}

\usepackage{amsmath}
\usepackage{graphicx}
\usepackage{url}
\usepackage{epsfig}
\usepackage[T1]{fontenc}
\usepackage[latin9]{inputenc}
\usepackage{multirow}

\newcommand{\tb}{\bar t}
\newcommand{\qb}{\bar q}
\newcommand{\ttbh}{ t \tb H}
\newcommand{\al}{\alpha}
\newcommand{\als}{\alpha_{\rm s}}
\newcommand{\shat}{\hat s}

\newcommand{\CA}{\mathswitch {C_{\mathrm{A}}}}
\newcommand{\sigh}{\hat \sigma}
\newcommand{\si}{\hat \sigma}
\newcommand{\nn}{\nonumber}

\newcommand{\tosv}{{\scriptscriptstyle \to}}

\def\beq{\begin{equation}}
\def\eeq{\end{equation}}
\def\bear{\begin{eqnarray}}
\def\eear{\end{eqnarray}}

\def\bet34{\beta_{kl}}
\def\CF{C_{\mathrm{F}}}
\def\CA{C_{\mathrm{A}}}

\title{Soft gluon resummation for associated $t \tb H$ production at the LHC}
\author[a]{Anna Kulesza,}
\author[b]{Leszek Motyka,}
\author[b]{Tomasz Stebel}
\author[a]{and Vincent Theeuwes}

\affiliation[a]{Institute for Theoretical Physics, WWU M\"unster, D-48149 M\"unster, Germany}
\affiliation[b]{Institute of Physics, Jagellonian University, S.\L{}ojasiewicza 11, 30-348 Krak\'ow, Poland}

\emailAdd{anna.kulesza@uni-muenster.de}
\emailAdd{leszekm@th.if.uj.edu.pl}
\emailAdd{tomasz.stebel@uj.edu.pl}
\emailAdd{vthee\_01@uni-muenster.de}

\abstract{We perform resummation of soft gluon corrections to the total cross section for the process $pp \to \ttbh$. The resummation is carried out at next-to-leading-logarithmic (NLL) accuracy using the Mellin space technique, extending its application to the class of $2 \to 3$ processes. We present an analytical result for the soft anomalous dimension for a hadronic production of two coloured massive particles in association with a colour singlet.  We discuss the impact of resummation on the numerical prediction for the associated Higgs boson production with top quarks at the LHC.}


\preprint{MS-TP-15-15}

\begin{document}
\maketitle

\section{Introduction}
\label{s:intro}
Establishing the properties of the Higgs boson discovered at the LHC in 2012~\cite{Aad:2012tfa, Chatrchyan:2012ufa}, in particular its couplings to the Standard Model (SM) particles, is one of the main tasks of the current LHC run.  Since the SM Higgs boson couples to fermions proportionally to their masses, the top-Higgs Yukawa coupling is expected to be  especially sensitive to the underlying physics.  A direct way to probe the strength of the coupling without making any assumptions regarding its nature is provided by the measurement of Higgs production rates in the $pp \to \ttbh$ process. Although the production cross section is low and the collision energy and the luminosity available so far have not been sufficient enough to measure a Higgs signal in Run 1~\cite{Aad:2014lma, Aad:2015gra, Khachatryan:2014qaa, Khachatryan:2015ila, Aad:2015iha},  such a measurement in Run 2 is eagerly awaited. Correspondingly, precision predictions for the $pp \to \ttbh$ production process are of great importance and a lot of effort has been invested in the recent years to improve the theoretical accuracy. 

The next-to-leading-order (NLO) QCD, i.e. ${\cal O}(\als^3\al)$ predictions are already known for some time ~\cite{Beenakker:2001rj, Reina:2001sf} and have been newly recalculated and matched to parton showers in~\cite{Hirschi:2011pa, Frederix:2011zi, Garzelli:2011vp, Hartanto:2015uka}. As of late, the mixed QCD-weak corrections~\cite{Frixione:2014qaa} and QCD-EW corrections~\cite{Yu:2014cka} of ${\cal O}(\als^2\al^2)$ are also available. Furthermore, the NLO QCD corrections to the hadronic $\ttbh$ production with top and antitop quarks decaying into bottom quarks and leptons have been recently obtained~\cite{Denner:2015yca}. Concurrently, new methods for a better measurement of the process have been proposed e.g. in~\cite{Plehn:2009rk} or in~\cite{Artoisenet:2013vfa}. In general, for the LHC collision energies of Run 2, the NLO QCD corrections are $\sim20$\%, whereas the size of the (electro)weak correction is more than ten times smaller. The scale uncertainty of the NLO QCD corrections is estimated to be $\sim10$\%~\cite{Beenakker:2001rj, Reina:2001sf, Dittmaier:2011ti}. 
While matching fixed-order predictions to parton showers pursued recently by many groups in such frameworks as aMC@NLO~\cite{Hirschi:2011pa, Frederix:2011zi, Alwall:2014hca}, POWHEG BOX~\cite{Alioli:2010xd, Garzelli:2011vp, Hartanto:2015uka} or SHERPA~\cite{Gleisberg:2008ta}  allows for a more accurate description of final state characteristics, it does not change the predictions for the overall production rates. An improvement in the accuracy with which these rates are known can only be achieved by calculating higher order corrections. However, calculations of the next-to-next-to-leading-order corrections are currently technically out of reach. It is nevertheless interesting to ask the question what is the size and the effect of certain classes of corrections of higher than NLO accuracy. In particular, we focus here on taking into account contributions from soft gluon emission to all orders in perturbation theory. The traditional (Mellin-space) resummation formalism which is applied in this type of calculations has been very well developed and copiously employed for description of the $2 \to 2$ type processes at the Born level. The universality of resummation concepts warrants their applications to scattering processes with many partons in the final state, as shown in a general analytical treatment developed for arbitrary number of partons~\cite{Bonciani:2003nt, Aybat:2006wq}. Recently, the soft gluon resummation technique in the soft collinear effective theory (SCET) framework was applied to $pp \to t\bar t W^{\pm}$ \cite{Li:2014ula}. So far, however, no calculations in the traditional resummation framework for processes involving $2\to 3$ scattering at the Born level have been performed.

In this paper we take the first step in this direction by developing the Mellin-space threshold resummation formalism at the next-to-leading-logarithmic (NLL) accuracy for the case of $2 \to 3$ processes with two coloured massive particles in the final state. We then apply this formalism in order to estimate the impact of soft gluon corrections on the predictions for the total $\ttbh$ production rate. In this particular case, the threshold region is reached when the square of the partonic center-of-mass (c.o.m.) energy, $\shat$, approaches $M=2 m_t +m_H$, where $m_t$ is the top quark mass and $m_H$ is the Higgs boson mass. In the threshold region, the cross section receives enhancement in the form of logarithmic corrections in $\beta = \sqrt {1-M^2/\shat}$. The quantity $\beta$ measures the distance from absolute production threshold and can be related to the maximal velocity of the $t \tb$ system. Additionally, in the threshold region the virtual QCD corrections are also enhanced due to Coulomb-type interactions between the two final state top quarks which become large when the top quark velocity in the $t \tb$  c.o.m. frame $\bet34 \to 0$ with $\bet34 =\sqrt{1- 4m_t^2/\shat_{kl}}$ and $\shat_{kl}=(p_t+p_{\tb})^2$. However, the contributions to the total cross section from the threshold region are strongly suppressed by the $\beta^4$ factor originating from the massive three particle phase space. Nevertheless, one expects that the threshold corrections can still have a non-negligible impact on the predictions.  

The associated production of a Higgs boson with a $t \tb$ pair involves four coloured partons at the Born level and as such is characterized by a non-trivial colour flow. The colour structure influences the contributions from wide-angle soft gluon emissions which have to be included at the NLL accuracy. The evolution of the colour exchange at NLL is governed by the one-loop soft anomalous dimension~\cite{BS, KS, KOSthr, KOScol, Bonciani:2003nt}. Starting from four coloured partons in the process, the soft anomalous dimension is a matrix and is known for heavy-quark ~\cite{KS,BCMNtop}, dijet~\cite{BS, KOSthr, KOScol} and supersymmetric particle production~\cite{Kulesza:2008jb, Beenakker:2009ha}, as well  as for the general case of $2 \to n$ QCD processes~\cite{Bonciani:2003nt, Aybat:2006wq}. Here we adopt the calculations of the soft anomalous dimension for the case of $2 \to 3$ processes with two coloured massive particles in the final state. 

\section{Resummation for $2 \to 3$ processes with two massive colored particles in the final state}
\label{s:theory}

The resummation of soft gluon corrections to the total cross section $\sigma_{pp \to \ttbh}$ is performed in Mellin space, where the Mellin moments are taken w.r.t. the variable $\rho = M^2/S$. At the partonic level, the Mellin moments for the process $ ij \to kl B$, where $i,j$ denote massless coloured partons, $k, l$ two massive quarks and $B$ a massive colour-singlet particle, is given by
\beq
\sigh_{ij \to kl B, N} (m_k, m_l, m_B, \mu_F^2, \mu_R^2) = \int_0^1 d \hat\rho \, \hat\rho^{N-1} \sigh_{ij \to kl B} (\hat \rho, m_k, m_l, m_B, \mu_F^2, \mu_R^2) 
\eeq
with $\hat \rho = 1- \beta^2$.

At LO, the $\ttbh$ production receives contributions from the $q \bar q$ and $gg$ channels. We analyze the colour structure of the underlying processes in the $s$-channel color bases, $\{ c_I^q\}$ and $\{c_I^g\}$, with \mbox{$c_{\bf 1}^q =  \delta^{\al_{i}\al_{j}} \delta^{\al_{k}\al_{l}},$} 
\mbox{$c_{\bf 8}^{q} = T^a_{\al_{i}\al_{j}} T^a_{\al_{k}\al_{l}},$}
\mbox{$c_{\bf 1}^{g} = \delta^{a_i a_j} \, \delta^{\al_k
  \al_l}, $}
\mbox{$c_{\bf 8S}^{g}=  T^b _{\alpha_l \alpha_k} d^{b a_i a_j} ,$}
\mbox{$c^{g}_{\bf 8A} = i T^b _{\alpha_l \alpha_k} f^{b a_i a_j} $}. 
In this basis the soft anomalous dimension matrix becomes diagonal in the production threshold limit~\cite{KS} and the the NLL resummed cross section in the
$N$-space has the form~\cite{KS, BCMNtop}
\beq
\label{eq:res:fact}
\sigh^{{\rm (res)}}_{ij\tosv kl B,N}\!\! =\!\! \sum_{I} \!\sigh^{(0)}_{ij\tosv
  klB,I,N}\, {C}_{ij\tosv kl B,I}\,
\Delta^i_{N+1} \Delta^j_{N+1} 
\Delta^{\rm (int)}_{ij\tosv kl B,I,N+1},
\eeq
where we suppress explicit dependence on the scales. 
The index $I$ in Eq.~(\ref{eq:res:fact}) distinguishes between contributions from
different colour channels. The colour-channel-dependent contributions
to the LO partonic cross sections in Mellin-moment space are
denoted by $\sigh^{(0)}_{ij\tosv klB,I,N}$. The radiative factors
$\Delta^i_{N}$ describe the effect of the soft gluon radiation
collinear to the initial state partons and are universal. Large-angle
soft gluon emission is accounted for by the factors $\Delta^{\rm
    (int)}_{ ij\tosv kl B,I,N}$ which depend on 
the partonic process under consideration and the colour configuration of
the participating particles. The expressions for the radiative factors in the $\overline{\mathrm{MS}}$ 
factorisation scheme read~(see e.g.~\cite{BCMNtop})              
\bear
\!&& \!\!\!\ln\! \Delta^{i}_N  =  \int_0^1\!\! dz \,\frac{z^{N-1}-1}{1-z}
\int_{\mu_{F}^2}^{ M^2(1-z)^2} \frac{dq^2}{q^2} A_i(\als(q^2))\, , \nn \\
 \!&& \!\!\!\ln \!\Delta^{\rm (int)}_{ij\tosv klB,I,N}\!\! = \!\!
 \int_0^1\!\!\! dz \frac{z^{N-1}-1}{1-z}  D_{ ij\tosv
   klB,I}(\als(M^2(1-z)^2)) . \nn
\eear
The coefficients $A_i,\,D_{ ij\tosv klB,I}$ are 
power series in the coupling constant $\als$,
\beq
A_i = (\frac{\als}{\pi}){A_i}^{(1)}+(\frac{\als}{\pi})^2
  {A_i}^{(2)}+ \dots, \ \  D_{ij\tosv klB,I} =
  (\frac{\als}{\pi}){D^{(1)}_{ ij\tosv klB,I}} + \dots
\eeq
The universal LL and NLL coefficients $A_i^{(1)}$, $A_i^{(2)}$ are
well known~\cite{KT,CET} and given by \mbox{$A_i^{(1)}= C_i$,} 
\mbox{$A_i^{(2)}=\frac{1}{2} \; C_i \left(\left( 
\frac{67}{18} - \frac{\pi^2}{6}\right) \CA- \frac{5}{9} n_f \right)$}  
with  \mbox{$C_g=\CA=3$,} and \mbox{$C_q=\CF=4/3$}.

The NLL coefficients $D_{ij\tosv klB,I}$ are obtained by taking the threshold limit $\shat \to M^2=(m_k+m_l +m_B)^2$  of the gauge-invariant soft anomalous dimension matrices $\Gamma^{ij \tosv klB}$. In this limit $\Gamma^{ij \tosv klB} = \frac{\als}{\pi} {\rm diag} (\gamma^{ij}_1,...)$ and $D_{ij\tosv klB,I} = 2 {\rm Re} (\gamma^{ij}_I)$. The calculations of $\Gamma^{ij \tosv klB}$ apply the methods developed in the heavy quark heavy quark pair-production~\cite{KS} to the process at hand, taking into account $2 \to 3$ kinematics, and yield
\beq
\label{eq:qqbargamma}
{\small
\Gamma ^{q\bar q\tosv k l B}=
{\alpha_s \over \pi} \,\left[
\begin{array}{cc}
\rule{0em}{2ex} - \CF (L_{\beta, kl}+1) & \frac{\CF}{\CA} \Omega_3 \\
\rule{0em}{2ex}  2 \Omega_3 & \frac{1}{2} [(\CA-2 \CF)(L_{\beta, kl}+1) +\CA \Lambda_3 +(8 \CF-3\CA)\Omega_3 ]\\
\end{array}
\right], 
}
\eeq 
\beq
\label{eq:gggamma}
{\small
\Gamma ^{gg\tosv k l B}=
{\alpha_s \over \pi} \,\left[
\begin{array}{ccc}
\rule{0em}{2ex} \Gamma_{11}^{gg} & 0 & \Omega_3\\
\rule{0em}{2ex} 0 & \Gamma_{22} ^{gg} &  \frac{N_c}{2} \Omega_3 \\
\rule{0em}{2ex}  2\Omega_3 & \frac{N_c^2-4}{2N_c} \Omega_3& \Gamma_{33}^{gg} \\
\end{array}
\right],
}
\eeq 
with 
\bear
&&\Gamma_{11}^{gg}  = - \CF (L_{\beta, kl}+1), \nn \\
&&\Gamma_{22}^{gg}  =\Gamma_{33}^{gg}= \frac{1}{2} ((\CA-2\CF)(L_{\beta, kl}+1) +\CA \Lambda_3), \nn 
\eear
where 
\bear
&&\Lambda_3= (T_1(m_k) +T_2(m_l)+U_1(m_l) +U_2(m_k))/2, \nn \\
&&\Omega_3= (T_1(m_k) +T_2(m_l)-U_1(m_l) -U_2(m_k))/2, \nn
\eear
and 
\bear
L_{\beta, kl} &=& \frac{\kappa^2  + \bet34^2}{2 \kappa \bet34}\left(\log\left( \frac{\kappa-\bet34}{\kappa+\bet34}\right) + i \pi\right),\\ 
T_i (m) &=& \frac{1}{2}\left( \ln((m^2- t_i)^2 / (m^2 \hat s )) -1 + i\pi \right),\\
U_i (m) &=& \frac{1}{2}\left( \ln((m^2-u_i)^2 / (m^2 \hat s )) -1 + i\pi \right),\\
\kappa &=& \sqrt{1-(m_k - m_l)^2 / s_{kl}},\qquad s_{kl}=(p_k+p_l)^2,\\
t_1=(p_i-p_k)^2,&&\quad t_2=(p_j-p_l)^2, \quad u_1=(p_i-p_l)^2, \quad u_2=(p_j-p_k)^2.
\eear
Eqs.~(\ref{eq:qqbargamma}),~(\ref{eq:gggamma}) reproduce the known results for  heavy quark-antiquark (squark-antisquark)  pair- production soft anomalous dimension~\cite{KS, Kulesza:2008jb} in the limit $p_B \to 0$. Also, our result for $\Gamma ^{q\bar q\tosv k l B}$ agrees with the result obtained in the SCET framework in~\cite{Li:2014ula}. It can be also explicitly seen that in the limit $\shat \to (2 m_t+m_H)^2$ the non-diagonal elements vanish and the diagonal elements give $D_{q \bar q\tosv klB,I}= \{0,-N_c \}$,  $D_{gg\tosv klB,I}=\{0,-N_c,-N_c\}$, which are the same coefficients as for the heavy-quark pair production $D_{ij \to kl}$. This confirms a simple physical intuition that the properties of the soft emission in the absolute threshold limit are only driven by the colour structure of the subprocesses and do not depend on the their kinematics.

The coefficients 
$$
{C}_{ij\tosv klB,I}= 1 + \frac{\als}{\pi} {C}^{(1)}_{ij\tosv klB,I}+ \dots
$$
contain all non-logarithmic contributions to the NLO cross section taken in the threshold limit. More specifically, these consist of Coulomb corrections, $N$-independent hard contributions from virtual
corrections and $N$-independent non-logarithmic contributions from soft emissions. 
Although formally the coefficients $C_{ij\tosv kl B,I}$ begin to contribute at NNLL accuracy, in our numerical studies of the $pp \to \ttbh$ process we consider both the case of $C_{ij\tosv kl B,I}=1$, i.e. with the first-order corrections to the coefficients neglected, as well as the case with these corrections included. In the latter case we treat the Coulomb corrections and the hard contributions additively, i.e.
$$
{C}_{ij\tosv klB,I}^{(1)}={C}_{ij\tosv klB,I}^{(1, \rm hard)}+{C}_{ij\tosv klB,I}^{(1, \rm Coul)} .
$$
For $k,l$ denoting massive quarks the Coulomb corrections are ${C}_{ij\tosv klB,{\bf 1}}^{(1, \rm Coul)} = \CF \pi^2 /(2 \beta_{kl})$ and  ${C}_{ij\tosv klB,{\bf 8}}^{(1, \rm Coul)} = (\CF -\CA/2) \pi^2 /(2 \beta_{kl})$. The additive treatment is consistent with NLL resummation and matching to NLO. We note that in general Coulomb corrections can also be resummed. A combined resummation of Coulomb and soft corrections is, however, beyond the scope of this paper. 

\section{Theoretical predictions for the $pp \to \ttbh$ process at NLO+NLL accuracy}
\label{s:results}

The resummation-improved NLO+NLL cross sections for the $pp \to \ttbh$ process are
obtained through matching the NLL resummed expressions with 
the full NLO cross sections
\bear
\label{hires}
&& \si^{\rm (NLO+NLL)}_{h_1 h_2 \tosv kl}(\rho, \mu_F^2, \mu_R^2)\! =\! 
\si^{\rm (NLO)}_{h_1 h_2 \tosv kl B}(\rho,\mu_F^2, \mu_R^2) +   \si^{\rm
  (res-exp)}_
{h_1 h_2 \tosv kl B}(\rho, \mu_F^2, \mu_R^2) \nn \\
&&\!\!\!\!\!\!\!\!\!{\rm with} \nn \\
&& \si^{\rm
  (res-exp)}_{h_1 h_2 \tosv kl B}  \! =   \sum_{i,j}\,
\int_{\cal C}\,\frac{dN}{2\pi
  i} \; \rho^{-N} f^{(N+1)} _{i/h{_1}} (\mu_F^2) \, f^{(N+1)} _{j/h_{2}} (\mu_F^2) \nn \\ 
&& \! \times\! \left[ 
\sigh^{\rm (res)}_{ij\tosv kl B,N} (\rho,\mu_F^2, \mu_R^2)
-  \sigh^{\rm (res)}_{ij\tosv kl B,N} (\rho,\mu_F^2, \mu_R^2)
{ \left. \right|}_{\scriptscriptstyle({NLO})}\! \right], 
\eear
where $\sigh^{\rm (res)}_{ij\tosv
  kl B,N}$ is given in Eq.~(\ref{eq:res:fact}) and  $ \sigh^{\rm
  (res)}_{ij\tosv kl B,N} \left. \right|_{\scriptscriptstyle({NLO})}$ represents its perturbative expansion truncated at NLO.
The moments of the parton 
distribution functions (pdf) $f_{i/h}(x, \mu^2_F)$ are 
defined in the standard way 
$f^{(N)}_{i/h} (\mu^2_F) \equiv \int_0^1 dx \, x^{N-1} f_{i/h}(x, \mu^2_F)$. 
The inverse Mellin transform (\ref{hires}) is evaluated numerically using 
a contour ${\cal C}$ in the complex-$N$ space according to the ``Minimal Prescription'' 
method developed in Ref.~\cite{Catani:1996yz}.

As mentioned in the previous section, the calculation of first-order contributions to the coefficients $C_{ij\tosv \ttbh,I}$ requires knowledge of the NLO real corrections in the threshold limit as well as virtual corrections. In our calculations we follow the methodology of~\cite{Beenakker:2011sf,Beenakker:2013mva}, where the case of two massive coloured particle in the final state was considered. We have explicitly checked that adding a massive colour singlet particle in the final state does not introduce any extra terms dependent on the mass of the added particle. Thus the $N$-space results for the pair-production process of two massive coloured particles are also applicable in our $2\to 3$ case. This way, the problem of calculating the $C_{ij\tosv \ttbh,I}^{(1)}$ coefficients reduces to calculation of virtual corrections to the process. We extract them numerically using the publicly available POWHEG implementation of the $\ttbh$ process~\cite{Hartanto:2015uka}, based on the calculations developed in~\cite{Reina:2001sf}. The results were then cross-checked using the standalone MadLoop implementation in aMC@NLO~\cite{Hirschi:2011pa}. Since the $q\qb$ channel receives only colour-octet contributions, the extracted value contributing to $C^{(1, {\rm hard})}_{q\qb \tosv t \tb H, {\bf 8}}$ is exact. In the $gg$ channel, however, both the singlet and octet production modes contribute. The implementation of the virtual corrections to $gg \to t \tb H$ in POWHEG and in aMC@NLO does not allow for their separate extraction in each colour channel. Instead, we extract the value which contributes to the coefficient $\bar{C}^{(1, {\rm hard})}_{gg \tosv t \tb H}$ averaged over colour channels and use the same value to further calculate $C^{(1,{\rm hard})}_{gg \tosv t \tb H, {\bf 1}}$ and $C^{(1,{\rm hard})}_{gg \tosv t \tb H, {\bf 8}}$. In order to measure the size of the error introduced by this procedure, we then rescale this value by the ratios of the corresponding colour-channel dependent and colour averaged coefficients found for $gg\to t\tb$ in~\cite{Czakon:2008cx}. The scale dependence of the $C_{ij\tosv kl B,I}^{(1)}$ can be fully deduced from renormalization group arguments, in the same way as for the full NLO result. We have checked that numerical results obtained with the procedure which we use to extract the values of the coefficients at $\mu_0=\mu_F=\mu_R$ show the same scale dependence as expected from exact analytical expressions.

In our phenomenological analysis we use $m_t=173$ GeV, $m_H=125$ GeV and choose the central scale $\mu_{F, 0} =\mu_{R, 0}= m_t +m_H/2$, in accordance with~\cite{Dittmaier:2011ti}. The NLO cross section is calculated using the aMC@NLO code~\cite{Alwall:2014hca}. In the implementation of the resummation formula, Eq.~\ref{eq:res:fact}, we numerically take a Mellin transform of the LO cross sections and the $C^{(1)}_{ij \tosv t \tb H,I}$ coefficient terms which are both calculated in the $x$ space. We perform the current analysis employing MMHT2014~\cite{Harland-Lang:2014zoa} pdfs and use the corresponding values of $\als$. Beside presenting the full result including non-zero $C^{(1)}_{ij \tosv t \tb H,I}$ coefficients, we also show the results with $C_{ij \tosv t \tb H,I}=1$.

\begin{figure}[h]
\centering
\includegraphics[width=0.45\textwidth]{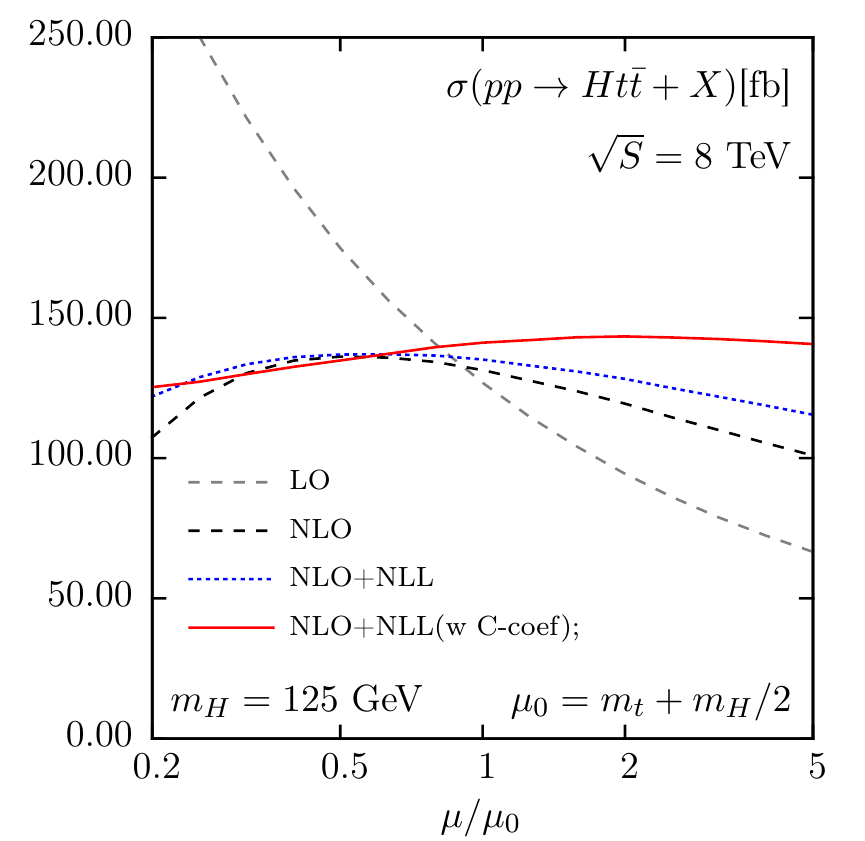}
\includegraphics[width=0.45\textwidth]{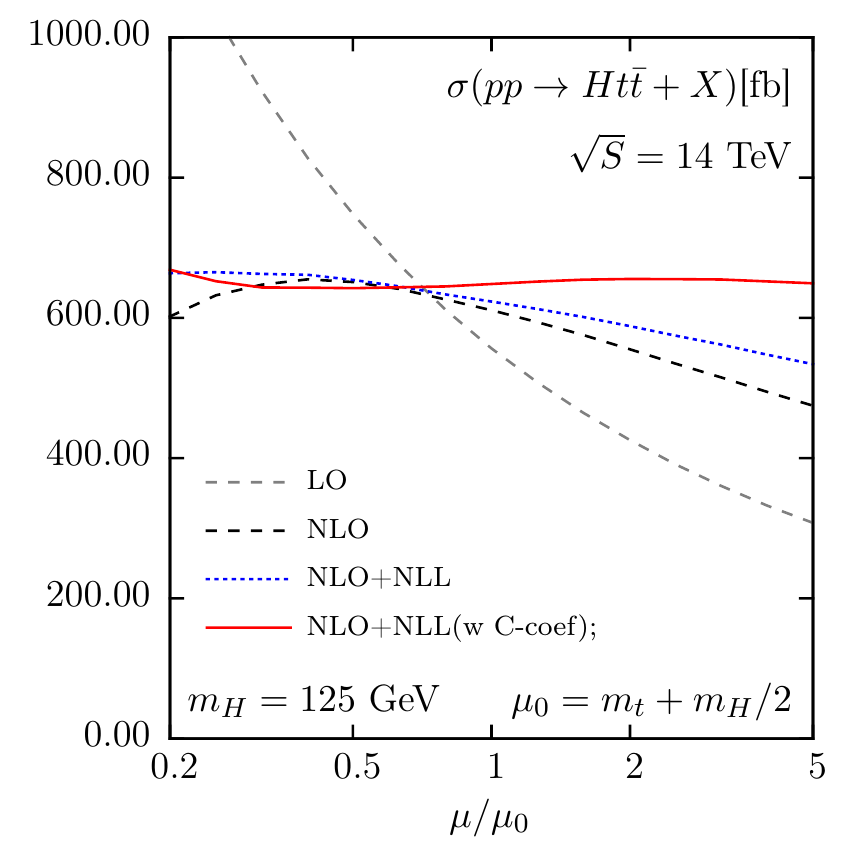}
\caption{Scale dependence of the LO, NLO and NLO+NLL cross sections at $\sqrt S=8$ and $\sqrt S=14$ TeV LHC collision energy. The results are obtained while simultaneously varying $\mu_F$ and $\mu_R$, $\mu=\mu_F=\mu_R$.} 
\label{f:scaledependence:sim}
\end{figure}

We begin our numerical study by analysing the scale dependence of the resummed total cross section for $pp \to \ttbh$ at $\sqrt S=8$ and 14 TeV, varying simultaneously the factorization and renormalization scales, $\mu_F$ and $\mu_R$. As demonstrated in Fig.~\ref{f:scaledependence:sim}, adding the soft gluon corrections stabilizes the dependence on $\mu=\mu_F=\mu_R$ of the NLO+NLL predictions with respect to NLO.  As an example, the central values and the scale error at $\sqrt S=8$ TeV changes from $132_{-9.3\%}^{+3.9\%}$ fb at NLO to $141_{-4.2\%}^{+1.4\%}$ fb at NLO+NLL (with $C^{(1)}_{ij \tosv \ttbh,I}$ coefficients included) and correspondingly, from $613_{-9.4\%}^{+6.2\%}$ fb to $650_{-1.2\%}^{+0.8\%}$ fb at $\sqrt S=14$ TeV. It is also clear from Fig.~\ref{f:scaledependence:sim} that the coefficients $C_{ij\tosv \ttbh}^{(1)}$ strongly impact the predictions, especially at higher scales. 

\begin{figure}[h]
\centering
\includegraphics[width=0.45\textwidth]{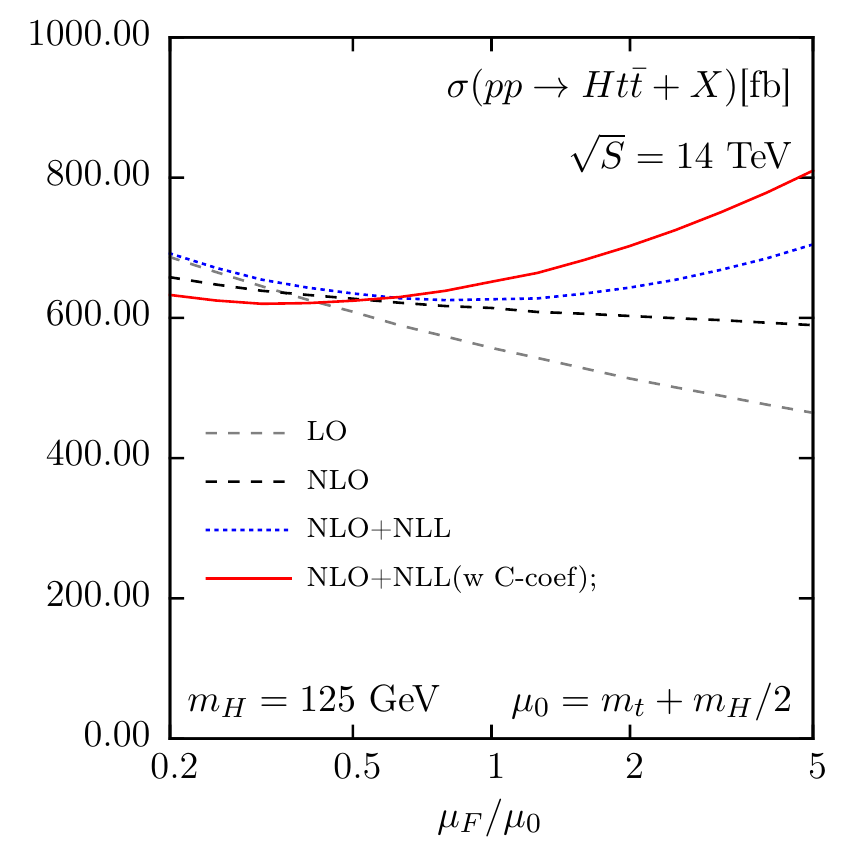}
\includegraphics[width=0.45\textwidth]{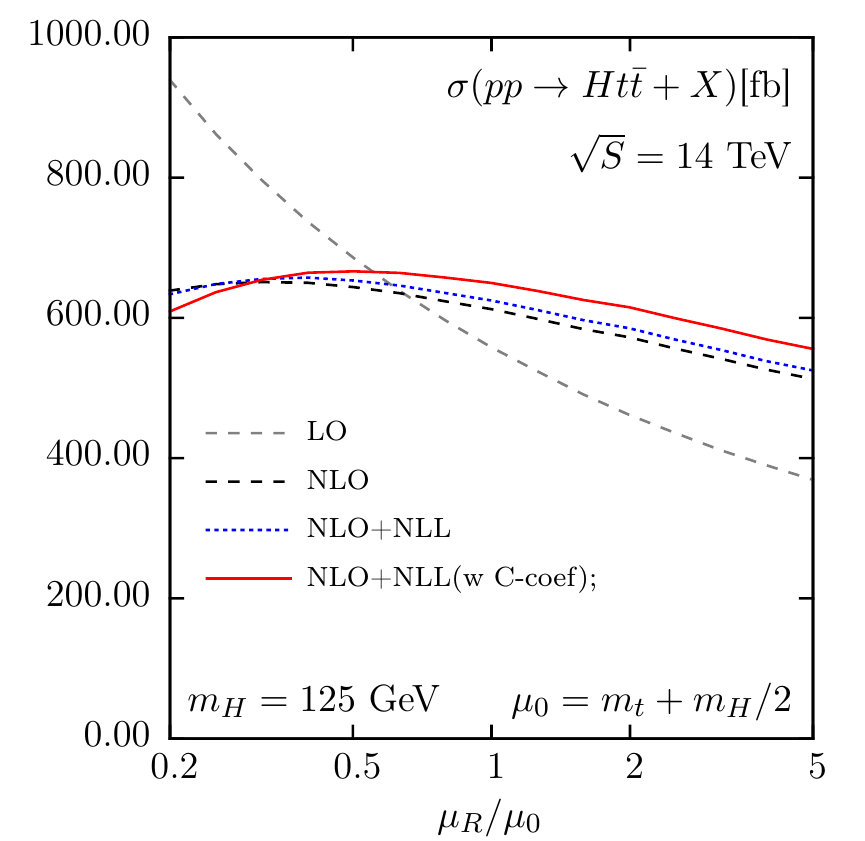}
\caption{Factorization and renormalization scale dependence of the LO, NLO and NLO+NLL cross sections at $\sqrt S=14$ TeV LHC collision energy. The results are obtained with $\mu_R=\mu_0$ for $\mu_F$ variation and and $\mu_F=\mu_0$ for $\mu_R$ variation.} 
\label{f:scaledependence:ind}
\end{figure}

In order to understand these effects better, in Fig.~\ref{f:scaledependence:ind} we analyse the dependence on the factorization and renormalization scale separately for the case study of $\sqrt S=14$ TeV. We observe that the weak scale dependence present when the scales are varied simultaneously is a result of the cancellations between renormalization and factorization scale dependencies. A similar effect of the opposite behaviour of the total cross section under $\mu_F$ and $\mu_R$ variations was previously shown for the total cross section for the inclusive Higgs production in the gluon-fusion process~\cite{Catani:2003zt}. The typical decrease of the cross section with increasing $\mu_R$ originates from running of $\als$. The behaviour under variation of the factorization scale, on the other hand, is related to the effect of scaling violation of pdfs at probed values of $x$. In this context, it is interesting to observe that the NLO+NLL predictions in Fig.~\ref{f:scaledependence:ind} show very little $\mu_F$ dependence around the central scale, in agreement with expectation of the factorization scale dependence in the resummed exponential and in the pdfs cancelling each other, here up to NLL. The relatively strong dependence on $\mu_F$ of the NLO+NLL predictions with non-zero $C^{(1)}_{ij \tosv t \tb H,I}$  can be then easily understood: the resummed expression will take into account higher order scale dependent terms which involve both $C^{(1)}_{ij \tosv t \tb H,I}$ and logarithms of $N$. These terms
 do not have their equivalent in the pdf evolution since the pdfs do not carry any process-specific information. Correspondingly, they are not cancelled and can lead to strong effects if the coefficients $C^{(1)}_{ij \tosv t \tb H,I}$ are numerically substantial. As these terms can only provide a part of the full scale dependence at higher orders, it is to be expected that their impact will be significantly modified when NNLO corrections are known.

%Given the arguments above, we conclude that the simultaneous $\mu_F$ and $\mu_R$ variation of the %resummed predictions cannot be a good measure of theoretical scale uncertainty. Therefore we choose to %estimate this uncertainty using the 7-point method, where the minimum and maximum values obtained with
Given the arguments above, we choose to estimate the theoretical uncertainty due to scale variation using the 7-point method, where the minimum and maximum values obtained with
 $(\mu_F/\mu_0, \mu_R/\mu_0) = (0.5,0.5), (0.5,1), (1,0.5), (1,1), (1,2), (2,1), (2,2)$ are considered. The effect of including NLL corrections is summarized in Table~\ref{t:results} for the LHC collision energy of 8, 13 and 14 TeV. The NLO+NLL predictions show a significant reduction of the scale uncertainty, compared to NLO results. The reduction of the positive and negative scale errors amounts to around 20-30\% of the NLO error for $\sqrt S=13, 14$ TeV and to around 25-35\% for $\sqrt S=8$ TeV. This general reduction trend is not sustained for the positive error after including the  $C^{(1)}_{ij \tosv t \tb H,I}$ coefficients. More specifically, the negative error is further slightly reduced, while the positive error is increased.  The origin of this increase can be traced back to the substantial dependence on $\mu_F$ of the resummed predictions with non-zero $C^{(1)}_{ij \tosv t \tb H,I}$ coefficients, manifesting itself at larger scales. However, even after the redistribution of the error between the positive and negative parts, the overall size of the scale error, corresponding to the size of the error bar, is reduced after resummation by around 7\% at 8 TeV and 10 (13)\% at 13 (14) TeV with respect to the NLO uncertainties.  The scale error of the predictions is still a few times larger than the pdf error,  cf. Table~\ref{t:results}. For simplicity, the pdf error shown in Table~\ref{t:results} is calculated for the NLO predictions, however adding the soft gluon correction can only minimally influence the value of the pdf error.

As expected on the basis of large phase-space suppression in the threshold regime, the predictions for total cross section at NLO+NLL are only moderately increased by 2-3\% w.r.t. the full NLO result. Introducing the coefficients $C^{(1)}_{ij \tosv t \tb H,I}$ leads to an increase in the $K$-factor of up to 6-7\%, indicating the importance of constant terms in the threshold limit. Since the impact of soft corrections is bigger for processes taking place closer to threshold the $K$-factor gets slightly higher for smaller collider energies. We also check the impact of our approximated treatment of  keeping parts of $C^{(1,{\rm hard})}_{gg\tosv t \tb H,{\bf 1}}$ and $C^{(1,{\rm hard})}_{gg\tosv t \tb H,{\bf 8}}$ coefficients coming from the virtual corrections equal to the colour channel averaged value, by rescaling at $\mu_F=\mu_R=\mu_0$ the averaged $\bar{C}^{(1,{\rm hard})}_{gg\tosv t \tb H}$ coefficient with ratios  $C^{(1,{\rm hard})}_{gg \tosv t \tb, I}/\bar{C}^{(1,{\rm hard})}_{gg \tosv t \tb}$ taken from~\cite{Czakon:2008cx}. The procedure is motivated by obvious similarities between the colour structures of the $pp\to t \tb$ and $pp \to \ttbh$ cross sections considered at threshold. We find that such rescaling of the hadronic $\ttbh$ cross section leads to a 3 per mille effect at 14 TeV, or a 5\% effect on the correction itself. Therefore we do not expect that the exact knowledge of the $C^{(1)}_{gg\tosv t \tb H,{\bf 1}}$ and $C^{(1)}_{gg\tosv t \tb H,{\bf 8}}$ coefficients will have a significant impact on the hadronic NLO+NLL predictions. However, we stress that because of the large phase-space suppression in the threshold regime the resummed results, while systematically taking into account a well defined class of correction, should not be used to estimate the size of the NNLO total cross section, by e.g. methods of expansion of the resummed exponential.

%\renewcommand{\arraystretch}{1.4}
%\begin{table}
%\begin{center}
%\begin{tabular}{|c|c|c|c|c|}
%\hline 
%$\sqrt{S}$ {[}TeV{]} & NLO {[}fb{]} & NLO+NLL {[}fb{]} & pdf uncertainty & $K$-factor\tabularnewline
%\hline 
%8 & $131_{-9.3\%}^{+3.9\%}$ & $135_{-5.7\%}^{+3.2\%}$ & $_{-2.7\%}^{+3.0\%}$ & 1.03\tabularnewline
%\hline 
%13 & $506_{-9.4\%}^{+5.9\%}$ & $517_{-6.6\%}^{+4.4\%}$ & $_{-2.3\%}^{+2.3\%}$ & 1.02\tabularnewline
%\hline 
%14 & $612_{-9.4\%}^{+6.2\%}$ & $624_{-6.6\%}^{+4.6\%}$ & $_{-2.2\%}^{+2.3\%}$ & 1.02\tabularnewline
%\hline 
%\end{tabular}
%\end{center}
%\label{tx:results}
%\caption{NLL+NLO and NLO total cross sections for $pp \to \ttbh$ for various LHC collision energies. The error ranges given together with the NLO and NLO+NLL results indicate the scale uncertainty.}
%\end{table}

\renewcommand{\arraystretch}{1.4}
\begin{table}
\begin{center}
\begin{tabular}{|c c c c c c c|}
\hline
$\sqrt{S}$ {[}TeV{]} & NLO {[}fb{]} & \multicolumn{2}{c}{NLO+NLL} & \multicolumn{2}{c}{NLO+NLL with $C$} &pdf error \tabularnewline
 & & Value {[}fb{]} & K-factor & Value {[}fb{]}  & K-factor & \tabularnewline
\hline 
8 & $132_{-9.3\%}^{+3.9\%}$ & $135_{-5.9\%}^{+3.0\%}$ & 1.03 &
 $141_{-4.6\%}^{+7.7\%}$&1.07 & $_{-2.7\%}^{+3.0\%}$ \tabularnewline
\hline 
13 & $506_{-9.4\%}^{+5.9\%}$ & $516_{-6.5\%}^{+4.6\%}$ & 1.02 &$537_{-5.5\%}^{+8.2\%}$ &1.06 & $_{-2.3\%}^{+2.3\%}$ \tabularnewline
\hline 
14 & $613_{-9.4\%}^{+6.2\%}$ & $625_{-6.7\%}^{+4.6\%}$ & 1.02 & $650_{-5.7\%}^{+7.9\%}$& 1.06 & $_{-2.2\%}^{+2.3\%}$ \tabularnewline
\hline 
\end{tabular}
\end{center}
\caption{NLO+NLL and NLO total cross sections for $pp \to \ttbh$ for various LHC collision energies. The error ranges given together with the NLO and NLO+NLL results indicate the scale uncertainty.}
\label{t:results}
\end{table}
%
%\sqrt{S}$ {[}TeV{]} & NLO {[}fb{]} & \multicolumn{2}{c}{NLO+NLL}  & \multicolumn{2}{c}{NLO+NLL with $C$} %&pdf error \tabularnewline

\section{Summary}

We have investigated the impact of the soft gluon emission effects on the total cross section for the process $pp\to t\tb H$ at the LHC. The resummation of soft gluon emission has been performed using the Mellin-moment resummation technique at the NLO+NLL accuracy. To the best of our knowledge, this is the first application of this method to a $2 \to 3$ process. Supplementing the NLO predictions with NLL corrections results in moderate modifications of the overall size of the total rates. The size of these modifications, as well as the size of the theoretical error due to scale variation is strongly influenced by the inclusion of the first-order hard matching coefficients into the resummation framework. The overall size of the theoretical scale error becomes smaller after resummation, albeit the reduction is relatively modest when the non-zero first-order hard matching coefficients are considered.

\section*{Acknowledgments}

This work has been supported in part by the DFG grant KU 3103/1. Support of the Polish National Science Centre grants no.\ DEC-2014/13/B/ST2/02486 is gratefully acknowledged. TS acknowledges support in the form of a scholarship of  Marian Smoluchowski Research Consortium Matter Energy Future from KNOW funding.

\bibliographystyle{JHEP}
\providecommand{\href}[2]{#2}\begingroup\raggedright\endgroup

\end{document}